\documentclass[12pt]{iopart}

\expandafter\let\csname equation*\endcsname\relax
\expandafter\let\csname endequation*\endcsname\relax
\usepackage{amsmath}

\usepackage{amssymb,amsthm}
\usepackage[dvipsnames]{xcolor}
\usepackage{xr}
\usepackage[titletoc,toc,title]{appendix}
\usepackage{hyperref}
\usepackage{graphicx,epstopdf,bm,amsfonts,dcolumn}
\hypersetup{colorlinks=true,linkcolor=blue,citecolor=red}
\usepackage{mathabx}
\usepackage{mathtools}

\renewcommand{\P}{{\mathbb P}}

\def\S{\mathbf{S}}

\begin{document}		

\title[Entropic Collapse and Extreme First-Passage Times]{Entropic Collapse and Extreme First-Passage Times in Discrete Ballistic Transport}

\author{Bhargav R. Karamched$^{1, 2, 3}$}
\address{$^1$ Department of Mathematics, Florida State University, Tallahassee, Florida 32306, USA}
\address{$^2$ Institute of Molecular Biophysics, Florida State University, Tallahassee, Florida 32306, USA}
\address{$^3$ Program in Neuroscience, Florida State University, Tallahassee, Florida 32306, USA}
\ead{bkaramched@fsu.edu}

\vspace{10pt}
\begin{indented}
\item[]\today
\end{indented}

\begin{abstract}
We investigate the extreme first-passage statistics of $N$ non-interacting random walkers on discrete, hierarchical networks. {By distinguishing between transport limited by escape from localized initial states (injection-limited) and transport limited by the extended network (bulk-limited), we identify a class of extreme value statistics that arises in geometries dominated by source traps (e.g., the Comet graph).} In this regime, the distribution of the minimum arrival time does not converge to any of the classical generalized extreme value distributions. Instead, it follows a discrete distribution with a {strict lower time bound} determined by the properties of the  hierarchical network. We analytically derive the asymptotic behavior of this class and validate our predictions against Monte Carlo simulations. Crucially, we identify the mechanism of ``entropic collapse" that destroys this scaling in bulk-dominated geometries like the Bethe lattice, where the phase space of delayed paths diverges with distance. This work establishes a geometry-encoding function that acts as a diagnostic tool for ascertaining whether or not a given graph is hierarchical.
\end{abstract}

\vspace{2pc}

\section{Introduction}
First-passage processes play a central role across statistical physics, probability theory, and the modeling of search phenomena~\cite{chou2014first,patie2004some,redner2001guide,hillen2025mean,d2026mean,newby2010quasi}. They arise naturally in problems ranging from chemical reaction kinetics and molecular transport~\cite{kolomeisky2007molecular} to neuronal signaling~\cite{sirovich2011spiking}, ecological foraging~\cite{byrne2012using}, and algorithmic search on networks~\cite{masuda2017random,noh2004random}. Of particular interest is the time required for a stochastic process to reach a specified target for the first time and how this time fluctuates across realizations.

In many applications, one is not concerned with a typical first-passage event but rather with the earliest arrival among many independent searchers. Such extreme first-passage times govern reaction onset, detection thresholds, and failure or success times in parallelized search strategies. For example, in biology, they are needed to understand the timescale for fertilization~\cite{saacke2000relationship} or infectious disease transfer~\cite{kumari2024first}. In fact, many events in cell biology are triggered by the fastest searcher in a group of searchers~\cite{dora2020active,schuss2019redundancy,patra2021level,basnayake2019fast,wong2026first,iyer2016first, allard2026tethered}. In decision dynamics, the extreme decider has been shown to strongly influence the rest of the members in a group~\cite{karamched2020bayesian,karamched2020heterogeneity,stickler2023impact}, even at the risk of revealing internal bias~\cite{linn2024fast}. In continuum settings, most notably Brownian motion, the statistics of the fastest particle have been studied extensively and are known to fall into classical extreme-value classes, typically of Gumbel or the Weibull type, governed by large deviations of diffusive trajectories~\cite{lawley2020distribution,lawley2020extreme,lawley2020probabilistic,lawley2020universal,linn2022extreme,lawley2021extreme,maclaurin2025extreme}.

However, a key structural feature of {many mathematical models of} real systems is that motion occurs in discrete space and discrete time. Examples include lattice-based transport models~\cite{tauber2014critical,bressloff2016model}, hopping dynamics on networks~\cite{lawley2020extreme}, random walks on graphs~\cite{masuda2017random}, and algorithmic searches constrained by adjacency rules~\cite{gramm2004automated}. {A natural question then concerns how the classical continuum theory pertains to these discrete models, or if an entirely new theory emerges.} In these systems, the geometry of the underlying space imposes a strict lower bound on the time required to reach a target. No trajectory can arrive earlier than the graph distance separating the source and the target. This introduces a {lower bound on the support of the first-passage-time distribution, fundamentally distinguishing discrete processes from their continuum counterparts. Moreover, these first-passage-time distributions exhibit an aggregation of probability mass at the lower bound.} 

Despite the ubiquity of discrete models, the extreme statistics of first-passage times in such settings remain comparatively less understood. Some recent studies have begun to explore extreme first-passage times for discrete random walks, demonstrating how randomly fluctuating environments drive the statistical variance of extreme arrivals~\cite{hass2024extreme}. However, the role of strict, deterministic graph topology in governing these extreme events remains an open question. 

In this work, we address this question by studying the extreme first-passage time (EFPT) for an ensemble of $N$ independent discrete-time random walkers on discrete graphs searching for a target.  {We ascertain a regime wherein the continuum theory coincides with the discrete theory--namely when $N$ is moderate so that the first arrival time is sufficiently far from the minimal time dictated by the topology of the graph. We further identify a phenomenon in the extreme value scaling regime we term `entropic collapse' where the EFPT emerges from a particle that locates the target in either the minimal time or the minimum possible delay value only. The EFPT distribution lacks a tail and consists of probability mass centered strictly at these two values. We reason that combinatorial complexity and `entropy' in discrete paths connecting the starting point to the target drives this phenomenon. 

{To see a nontrivial EFPT distribution, we therefore investigate discrete graphs whose path entropy is localized to a particular subgraph. We term these graphs discrete hierarchical graphs.} These graphs consist of a head consisting of a {connected graph, which we also call a \textit{source trap}, that feeds into a long tail of ballistic, unidirectional motion (see Figure~\ref{fig:dhg}). Here, particles must navigate this source trap and experience stochastic delays before successfully entering the unidirectional tail. Thus, particle dynamics along the graph consists of two distinct regimes: (1) an injection regime, where particles are stuck in the source trap and (2) a transport regime, where particles move ballistically. When the injection regime dominates the time required to reach a target, we call such processes \textit{injection-limited.} When the transport regime dominates the time to reach a target, we call such processes \textit{bulk-limited}.} 

\begin{figure}
    \centering
    \includegraphics[width=\linewidth]{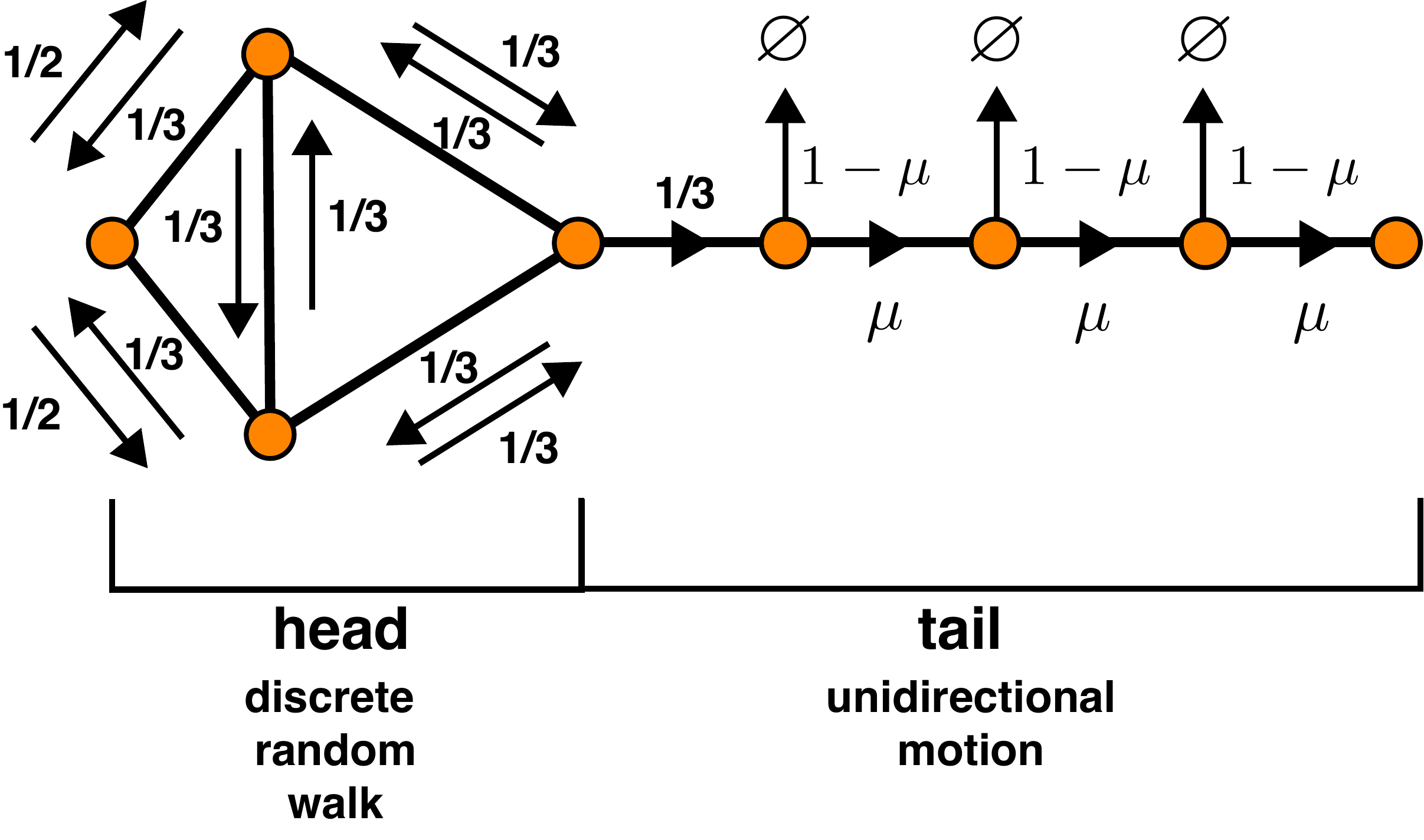}
    \caption{Example of a discrete hierarchical graph. Within the head, a particle undergoes an unbiased random walk, transitioning to adjacent nodes with a probability of $k^{-1}$, where $k$ is the degree of the current node. Upon entering the tail, the particle undergoes unidirectional motion, advancing with probability $\mu$ and decaying with probability $1-\mu$ at each time step.}
    \label{fig:dhg}
\end{figure}

We focus on the earliest arrival time,
$$
T_N = \min\{\tau_1,\tau_2,...,\tau_N\},
$$
where each $\tau_i$ denotes the first-passage time of an individual particle moving on the discrete hierarchical network. Our analysis reveals that, in contrast to continuum diffusion, the extreme statistics in discrete geometries are dominated by ballistic, shortest-path trajectories, with longer paths contributing only entropic corrections.

We identify a scaling regime in which the number of walkers $N$ grows while the probability $p_d$ of a single walker reaching the target in the minimal number of steps, $d$, becomes small, such that $Np_d$ remains finite. Equivalently, we consider the regime where $N$ grows and $d$ grows such that $p_d$ becomes small and $Np_d$ remains finite. In this limit, the earliest arrival behaves as a Poisson process of rare, nearly deterministic trajectories. The resulting extreme-value distribution exhibits a lower bound at the graph distance $d$ and takes an exponential form that is not captured by classical Gumbel, Weibull, or Fr\'{e}chet laws.

A central object in our theory is a geometry-dependent function $F(k)$, which quantifies the cumulative contribution of paths that arrive within $k$ additional steps beyond the shortest possible time. This function encodes the entropic suppression of detours away from geodesic paths and provides a transparent link between graph structure and extreme statistics. When this entropic suppression is sufficiently strong, the extreme behavior is governed entirely by near-geodesic trajectories.

Our framework highlights a sharp distinction between different classes of discrete geometries. In some graphs, such as finite-dimensional lattices or networks with constrained branching, shortest paths remain dominant even as the source-to-target distance grows. In others, including trees with exponential volume growth, the proliferation of near-geodesic excursions leads to a breakdown of the ballistic scaling regime. Rather than being a limitation, this distinction delineates different classes of extreme first-passage behavior determined by geometric properties. We develop a criterion based on entropic cost that delineates between injection-limited processes and bulk-limited processes: The geometry of a graph determines whether extreme first-passage statistics are controlled by the source trap or by the bulk, and this is precisely diagnosed by whether $F(k)$ is independent of $d$.

{Our framework further clarifies the structure of EFPTs when a minimal time imputed by the graph geometry is considered. While recent foundational work has thoroughly investigated how minimal time bounds shape extreme-value frameworks in specific network and diffusion contexts~\cite{lawley2020extreme,linn2022extreme}, the role of strict, deterministic graph topology in governing these extreme events---specifically regarding the combinatorial onset of entropic collapse---remains comparatively less understood.}

{By retaining the discrete nature of space and time, our results capture qualitative features—such as combinatorial entropic collapse—that are specific to discrete topologies and crucial for understanding extreme events in discrete models.}

\section{{Motivating Example: the 1D Lattice}}
{Consider an ensemble of $N$ particles each performing a discrete symmetric random walk on an infinite one-dimensional lattice beginning at the same site. That is, in a given time step, each particle moves to the right with probability 1/2 or to the left with probability 1/2. Assume a target exists a distance $d$ away. What is the extreme first passage time for the first of these particles to locate the target?}

\begin{figure}
    \centering
    \includegraphics[width=\linewidth]{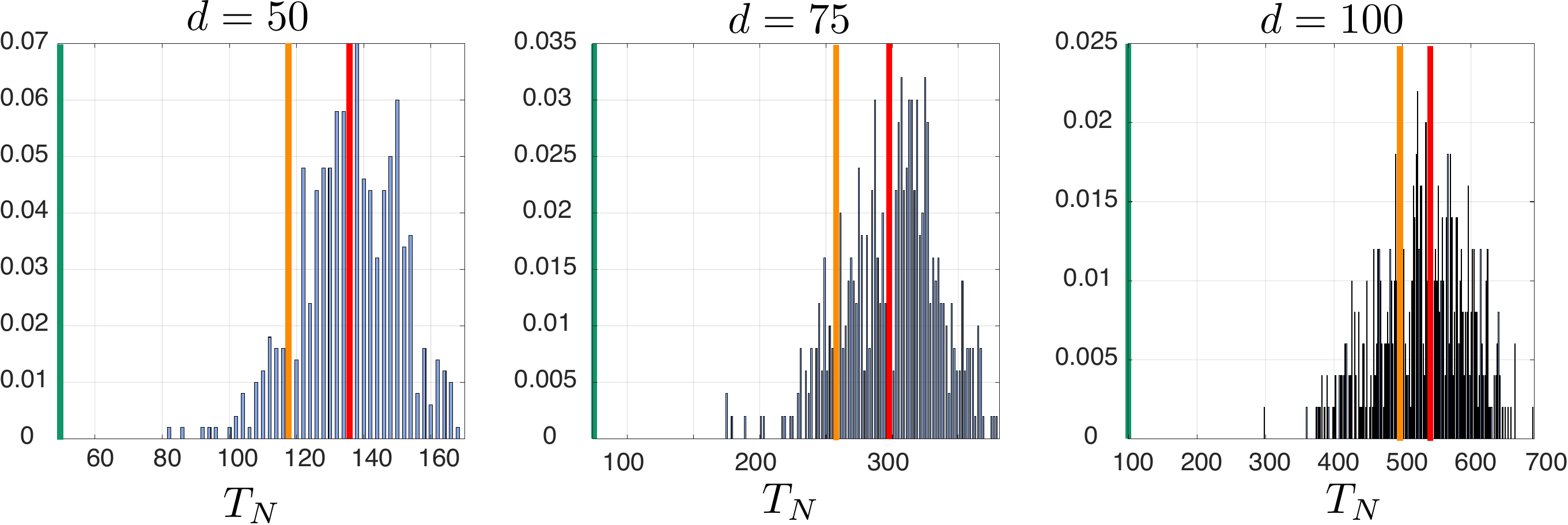}
    \caption{{Histograms of extreme first passage time, $T_N$, with $N = 50000$ particles performing a discrete symmetric random walk on a 1D lattice for different target distances. Green line indicates distance. Orange line indicates $\langle T_N \rangle$ as predicted by the continuum theory, Eq.~\eqref{eq:continuum}. Red line is simulated value of $\langle T_N \rangle$ over 500 trials.}}
    \label{fig:continuum_comparison}
\end{figure}

{The continuum theory of Brownian motion gives us the elegant expression for the mean EFPT~\cite{lawley2020universal,lawley2020distribution,schuss2019redundancy,karamched2020heterogeneity}
\begin{equation}
    \langle T_N \rangle = \frac{d^2}{4D\ln{N}},
    \label{eq:continuum}
\end{equation}
obtained as an asymptotic approximation in the limit $N \to \infty$. Remarkably, Eq.~\eqref{eq:continuum} predicts the mean EFPT for the discrete case when $N$ is finite and $d$ is large so that the discrete system is near the diffusion limit (see Figure~\ref{fig:continuum_comparison}). In this moderate $N$ regime, the fastest particles arrive at times still far greater than the absolute minimum distance $d$. Because these extreme arrivals do not interact with the rigid lower bound imputed by the lattice, the Central Limit Theorem holds, and continuous diffusion accurately approximates the discrete dynamics~\cite{bressloff2013stochastic,van1992stochastic,gardiner2009stochastic}. But then the question remains: how does $T_N$ scale as $N\to \infty$ in the discrete setting?}

{For a fixed $d$, taking $N \to \infty$ yields the trivial result that mean EFPT is $d$ and in fact the distribution is a Dirac mass centered at $T_N = d$. For the limiting behavior of the system to be nontrivial, we need to take $N \to \infty$ and $d \to \infty$ so that $Np_d = \lambda \in (0,\infty)$, where $p_d$ is the probability that a single walker takes the shortest path to the target.} 

{In Figure~\ref{fig:continuum_collapse}, we show the distributions for $T_N$ emerging from scaling $N$ and $d$ as described above. Interestingly, as $d$ increases, the tail of the distribution vanishes. We attribute this to the combinatorial complexity in the paths that connect the starting site to the target. There is exactly one path that takes a particle from the starting site to the target in $d$ steps. Because $N$ is large, with high probability the first arrival follows that path. But as $d$ is increased, the number of paths that connect the starting site to the target that are delayed by the minimum amount increases as well. For the 1D lattice, the minimal delay amount is two: a particle must move away from the target and then correct its mistake to reach the target quickly. Because the 1D lattice is translationally invariant, a delayed path can be generated by taking a 'wrong' step at any vertex along the route. If $d$ is large, there are so many minimally delayed paths that the first arrival, if it did not take the shortest path, is forced to take a minimally delayed path. The combinatorial complexity causes the mass across the tail of the distribution to `collapse' to $d + 2$.}

{This `path entropy' permeates the space and forces the emergence of these non-classical extreme value distributions: the collapsed distribution does not resemble a Gumbel, Weibull, or a Fr\`{e}chet. We term this obliteration of the distribution tail as `entropic collapse'.}

\begin{figure}
    \centering
    \includegraphics[width=\linewidth]{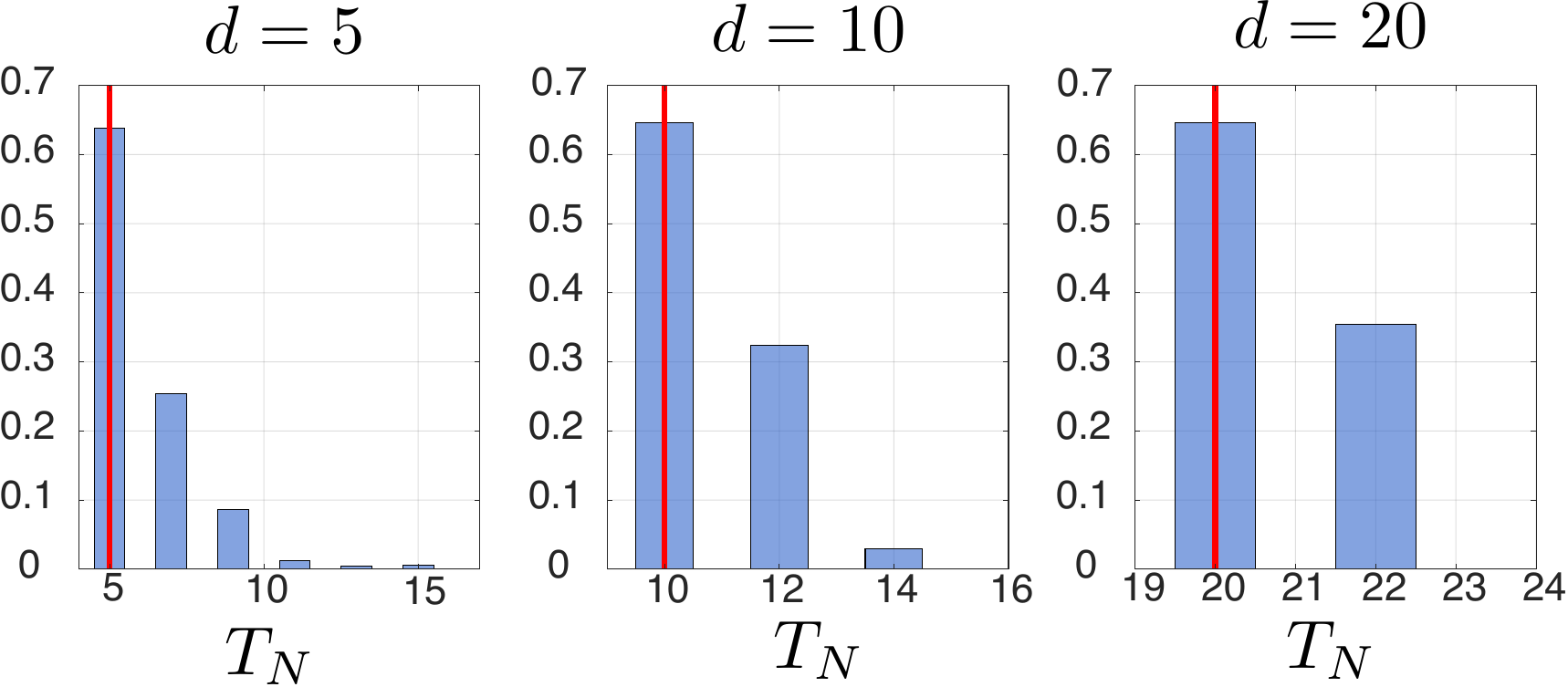}
    \caption{{Histograms of extreme first passage time, $T_N$, with $N = \lambda/p_d$ particles performing a discrete symmetric random walk on a 1D lattice for different target distances. Red line indicates distance. Shown histogram is obtained by performgin 500 simulations. Here, $\lambda = 1$ and $p_d = (1/2)^d$.}}
    \label{fig:continuum_collapse}
\end{figure}

{ The 1D lattice is therefore a bulk-limited geometry: the combinatorial growth of delayed paths occurs throughout the transport channel rather than remaining localized. What graphs do not experience entropic collapse in the joint scaling limit? This is the motivating question for this work. Presumably, graphs that localize path entropy to a subgraph will not experience this collapse. As such, we investigate EFPTs on discrete hierarchical graphs.}

\section{General Framework and Main Result}
\label{sec:framework}

{Consider a discrete hierarchical graph $\mathcal{G}$ with finite or countably infinite nodes. That is, let $\mathcal{G}$ be a graph that is divided into subgraphs $\mathcal{G}= \mathcal{H} \cup \mathcal{T}$, where $\mathcal{H}$ is a finite, connected network (the head) and $\mathcal{T}$ is a unidirectional network (the tail). The subgraph $\mathcal{H}$ contains a unique node $g_{out}\in \mathcal{H}$ which extends an edge to a single node on $\mathcal{T}$. The tail $\mathcal{T}$ contains a node $g_t$ that we call the target.}

Consider an ensemble of $N$ particles all located at the same site $g_0 \in \mathcal{H}$ initially. We assume the following particle dynamics: in $\mathcal{H}$ a particle performs a discrete time random walk, moving with equal probability to any of its neighbors. In $\mathcal{T}$, a particle moves unidirectionally from node $i \to i+1$ with probability $\mu$ in a time step. Thus, along $\mathcal{T}$, each particle has a probability $1-\mu$ of decaying. Physically, these graphs represent reaction dynamics within $\mathcal{H}$ coupled to ballistic transport dynamics along $\mathcal{T}$.

{We are interested in understanding the extreme first-passage time statistics of this system as the distance to the target becomes macroscopically large. To make this ``large graph" limit mathematically rigorous, we consider a sequence of triples $\{(\mathcal{G}_n, s_n, t_n)\}_{n=1}^\infty$, where $\mathcal{G}_n = \mathcal{H}_n \cup \mathcal{T}_n$ is a discrete hierarchical graph, $s_n \in \mathcal{H}_n$ is the starting site, and $t_n \in \mathcal{T}_n$ is the target site. We of course assume that $s_n \neq t_n$ because in that case the distance between the nodes is zero, which is not a particularly interesting case.}

{ Let $\delta_n(a,b)$ describe the shortest-path distance between nodes $a$ and $b$ induced by allowed moves on $\mathcal{G}_n$. The distance from starting point to the target on this sequence of triplets is $d_n = \delta_n(s_n,t_n) \in \mathbb{N}$. We define the large graph limit as the case when $d_n \to \infty$ as $n \to \infty$. }

We now describe how extreme first passage times tie in with these discrete hierarchical graphs. Let $\tau_i$ represent the first-passage time for particle $i$ to reach $t_n$ from $s_n$. We wish to understand the asymptotic distribution of the earliest arrival, $T_N = \min\{\tau_1,\tau_2,...,\tau_N\}$. In stark contrast to classical Brownian processes, $\mathbb{P}(\tau_i < d_n) = 0$. The minimum number of steps required to reach the target equals the distance induced by the metric $\delta_n$, imputing a lower bound to the extreme distribution. {For brevity of notation in the ensuing derivations, we will suppress the sequence index $n$ unless explicitly stated.  We will refer to the distance as $d$, with the understanding that the limit $d \to \infty$ is taken along the formal sequence of triples.}

{If we define $p_j$ as the probability that single walker reaches the target in $j$ steps, $p_j = \mathbb{P}(\tau = j)$, then the probability that a single walker reaches the target in the minimum number of steps, $d$, is $p_d$. Consequently, $\mathbb{P}(\tau > d) = 1-p_d$. Because the particles evolve independently, we therefore have that 
\begin{equation}
\mathbb{P}(T_N = d) = 1- (1-p_d)^N \quad \quad \mathbb{P}(T_N > d) = (1-p_d)^N.
\label{eq:TN}
\end{equation}}
{In general, by independence, we have that
\begin{equation}
\mathbb{P}(T_N>t)  = S(t)^N,
\end{equation}
where $S(t) \equiv \mathbb{P}(\tau > t)$ is the survival probability of a single particle. We can use the survival probabilities to express the mean extreme first passage time
\begin{equation}
\langle T_N\rangle = \sum_{k=0}^\infty \mathbb{P}(\tau > k)^N = \sum_{k = 0}^\infty S(k)^N = d + \sum_{k=d}^\infty S(k)^N = d + (1-p_d)^N + \sum_{k=d+1}^\infty S(k)^N.
\label{eq:meanTN}
\end{equation}
The third equality in Eq.~\eqref{eq:meanTN} follows from the fact that $S(k) = 1$ for $k = 0,1,2,...,d-1$ and the fourth equality follows from Eq.~\eqref{eq:TN}.}

{We will show that $\langle T_N \rangle$ and the distribution for $T_N$ have an unexpected structure in the case that contributions to $\langle T_N \rangle$ coming from arrival times exceeding $d$ are bounded on a graph. By bounded, we mean that there exists a finite, nondecreasing function $F(k)$ with $F(0) = 1$ such that 
\begin{equation}
\sum_{j=0}^k p_{d+j}
=
p_{d} \, F(k) + o(p_{d}).
\label{eq:entropy}
\end{equation}}
This condition expresses an entropic penalty for detours. It prevents contributions from paths longer than the shortest paths from dominating the extreme statistics. {We say arrival probabilities $p_j$ that satisfy Eq~\eqref{eq:entropy} are entropically bounded.} 

{To understand the statistics of the extreme first passage time in the large graph limit, we need to consider a scaling regime in which shortest-path arrivals are rare events but the expected number of such arrivals remains finite, $N\to \infty$ and $p_d \to 0$ such that $Np_d \to \lambda \in (0,\infty)$. Such a scaling is necessary because if $p_d \to 0$ but $N$ is finite, then $\mathbb{P}(T_N =d) = 0$. Conversely, if $N \to \infty$ but $p_d$ is nonzero, $\mathbb{P}(T_N = d) = 1$. The joint limit encoded in $Np_d \to \lambda$ corresponds to an extreme-value scaling regime. Thus, physically, $\lambda$, represents the expected number of particles to reach the target in the shortest path of $d$ steps.}

{From Eq.~\eqref{eq:meanTN}, we have 
\begin{equation}
    \langle T_N\rangle = d + \sum_{k=0}^\infty S(d+k)^N
    \end{equation}
    Since
    \begin{equation}
    S(d+k) =  1 - \sum_{j=0}^k p_{d+j},
    \end{equation}
    if we assume entropic boundedness, we have
    \begin{equation}
    S(d+k)^N = \left(1 - \sum_{j=0}^k p_{d+j}\right)^N = \left(1 - p_dF(k) + o(p_d)\right)^N
    \end{equation}
    Our assumption in Eq.~\eqref{eq:entropy} allows us to write that, as $p_d \to 0$, 
    \begin{align}
    S(d+k)^N &= \left(1 - p_d F(k) + o(p_d)\right)^N
\approx \exp\!\left(-N p_d F(k)\right)\,(1+o(1)).
\label{eq:survival}
    \end{align}
   The $o(1)$ appears at the end since $Np_d \to \lambda \in (0,\infty)$.
    Thus,
    \begin{equation}
    \langle T_N\rangle = d + \sum_{k=0}^\infty \exp\left(-Np_dF(k)\right) + o(1).
    \label{eq:meanTN2}
    \end{equation}}

   \noindent For the distribution, recall that for a single walker, we have
   \begin{equation}
   \mathbb{P}(\tau > d + k) = 1 - \sum_{j=0}^k p_{d+j},
   \end{equation}
   and, by independence,
   \begin{equation}
   \mathbb{P}(T_N > d+k) = \left(1 - \sum_{j=0}^k p_{d+j}\right)^N = \left(1 - p_dF(k) + o(p_d)\right)^N.
   \end{equation}
  If we now invoke our extreme value scaling $N \to \infty$, $p_d \to 0$, with $Np_d \to \lambda$, we have
   \begin{equation}
   \mathbb{P}(T_N>d+k) \approx \left(1-\frac{\lambda}{N}F(k) + o(p_d)\right)^N \to \exp\left(-\lambda F(k)\right) \text{  as  } N\to \infty
   \label{eq:distTN}
   \end{equation}
   We note that this assumes that single walker success is very rare, many walkers exist, and the expected number of shortest path successes ($\lambda$) stays finite.
\\

Importantly, Eq.~\eqref{eq:distTN} is not the Weibull distribution or any of the distributions standard in classical extreme value theory (e.g., the Gumbel and the Fr\'{e}chet distributions). It is an extreme-value distribution with a probability mass accumulating lower bound. A physical interpretation of this limit is that the first arrival of a particle behaves like a Poisson process of ballistic trajectories. 

An immediate corollary of this result is that one can obtain asymptotic expressions for all the moments of $T_N$:
\begin{align}
    \left\langle(T_N - d)^m\right\rangle &= \sum_{k=0}^\infty \left[(k+1)^m - k^m\right]\mathbb{P}(T_N - d > k)\nonumber\\ 
    &\approx \sum_{k=0}^\infty \left[(k+1)^m - k^m\right]\exp\left(-\lambda F(k)\right)
    \label{eq:moments}
\end{align}

We note that while the exponential form in Eq.~\eqref{eq:survival} accurately captures the asymptotic scaling behavior in the relevant regime (where $k \sim \ln{N}$~\cite{flajolet2006ubiquitous}), the exact survival probabilities in Eq. ~\eqref{eq:meanTN} should be used when evaluating the infinite sum to ensure convergence in the far tail, where the exponential approximation saturates to a non-zero constant. Similarly, Eq.~\eqref{eq:meanTN2} describes the \textit{body} of the distribution. Using it as an exact distribution leads to divergence issues. The expression for $\langle T_N\rangle$ without the exponential approximation,
$$
\sum_{k=1}^\infty\left(1 - p_d F(k)\right)^N
$$
converges as a geometric series, but the exponential approximation must be truncated.

A key point here is that the structure of the function $F(k)$ is a diagnostic in ascertaining entropic boundedness of a graph (see Section 3 below): if $F(k)$ is independent of $d$, the arrival probabilities on the graph are entropically bounded and the exponential form of the extreme first passage time distribution in Eq.~\eqref{eq:distTN} holds.

In the next section, we apply these expressions for $\langle T_N\rangle$ and $\mathbb{P}(T_N > d+k)$ to concrete examples.

\section{Examples}
\label{sec:examples}
We apply our framework to three geometries: (1) a Comet network, for which our theory holds, (2) the Bethe lattice, for which our theory fails (Figure~\ref{fig:neighborhood}), and (3) a Braided Tail, for which our theory holds. In the Comet network, the dynamics along the tail have no impact upon the shape of $F(k)$, but the tail is essential for our result to hold. In a Braided Tail, the multiple tails affect the shape of $F(k)$, imputing multiple modes into the distribution for $T_N$. {We explicitly define the sequence of triples for each geometry to demonstrate exactly how the macroscopic limit $d \to \infty$ is taken.}

\begin{figure}
    \centering
    \includegraphics[width=\textwidth]{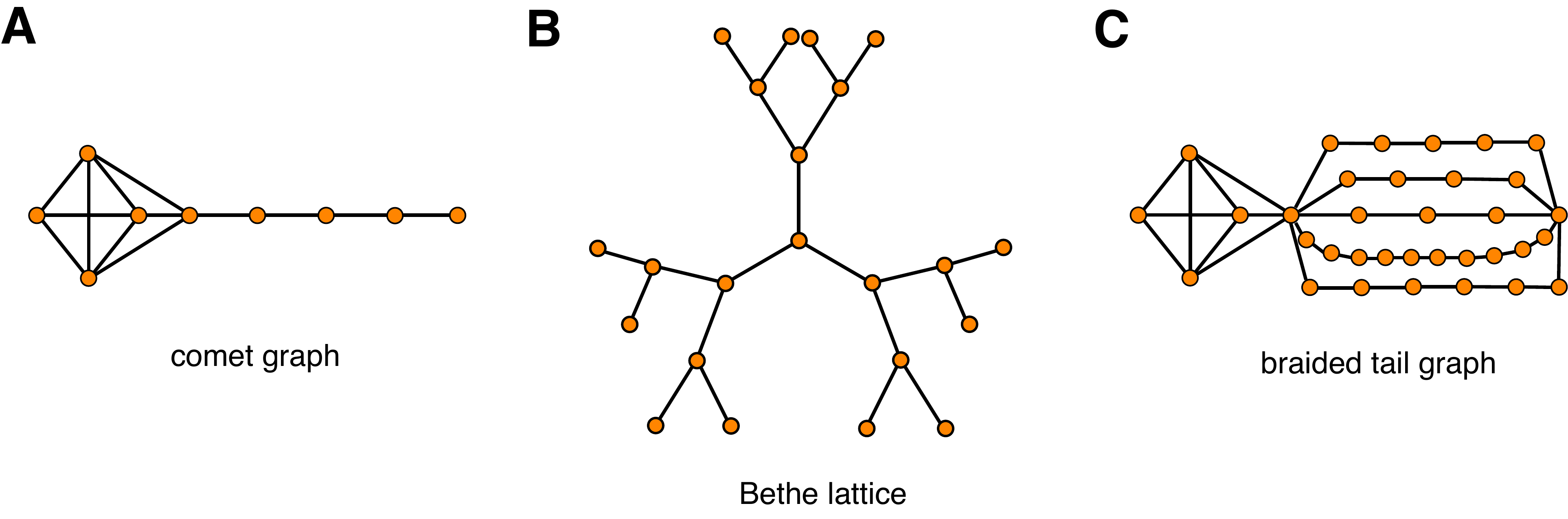}
  \caption{{Graphs to which we apply our theory. (A) an example of a comet graph whose head, $\mathcal{H}$, is formed by a connected graph of four nodes and whose tail, $\mathcal{T}$, is formed by a unidirectional network of four nodes. Our theory works for this graph (see text). (B) an example of a truncated Bethe lattice at three generations with a coordination number of 3. Our theory fails for this graph because arrival probabilities lack entropic boundedness. (C) a braided tail graph with $\mathcal{H}$ is formed by a connected graph of four nodes and $\mathcal{T}$ formed with multiple lanes of different lengths.} }
    \label{fig:neighborhood}
\end{figure}

\subsection{Directed Random Walk with a Source Trap}

We first consider the ``Comet Graph,'' which separates the domain into a finite ``Head'' $\mathcal{H}$ (a localized trap) and a ``Tail'' $\mathcal{T}$ (a directed transport channel). 

{To evaluate the asymptotic limit, we define the sequence of triples $\{(\mathcal{G}_n, s_n, t_n)\}_{n=1}^\infty$. Let the head $\mathcal{H}$ remain structurally fixed for all $n$. Let the tail $\mathcal{T}_n$ consist of a directed chain of length $L_n = n$. We fix the start site $s_n = g_0 \in \mathcal{H}$ for all $n$. There exists a designated exit node $g_{out} \in \mathcal{H}$ connecting to the first node of $\mathcal{T}_n$. The target $t_n$ is defined as the terminal node of the tail $\mathcal{T}_n$. Thus, the shortest-path distance is $d_n = d_{\mathcal{H}} + n$, which diverges as $n \to \infty$.} Because $L_n$ grows but $\mathcal{H}$ is fixed, all delayed excursions must occur strictly within the head. 


Dropping the subscript $n$ for brevity, the probability of reaching the target in the minimum number of steps $d$ is:
\begin{equation}
    p_d = \pi_{d_{\mathcal{H}}}(g_0, g_{out}) \cdot \mu^L,
\end{equation}
where $\pi_m(u,v)$ denotes the probability of transitioning from $u$ to $v$ in exactly $m$ steps within $\mathcal{H}$ and $\mu$ is the probability of the particle being removed on a given time step along $\mathcal{T}$ (as discussed in Section 2). Any delayed arrival of time $d+k$ implies the particle spent an additional $k$ steps wandering inside $\mathcal{H}$:
\begin{equation}
    p_{d+k} = \pi_{d_{\mathcal{H}}+k}(g_0, g_{out}) \cdot \mu^L.
\end{equation}

Testing the entropic boundedness condition along our sequence:
\begin{equation}
    \frac{p_{d+k}}{p_d} = \frac{\pi_{d_{\mathcal{H}}+k}(g_0, g_{out}) \mu^L}{\pi_{d_{\mathcal{H}}}(g_0, g_{out}) \mu^L} = \frac{\pi_{d_{\mathcal{H}}+k}(g_0, g_{out})}{\pi_{d_{\mathcal{H}}}(g_0, g_{out})}.
\end{equation}
Because the sequence scales only the directed tail, the $\mu^L$ terms cancel. The ratio is strictly independent of $d$. Thus, $F(k)$ converges to a constant dependent only on the fixed local trap geometry, perfectly capturing the entropic boundedness of the graph.

\subsubsection{The Leaky Loop Model}

\begin{figure}
    \centering
    \includegraphics[width=\linewidth]{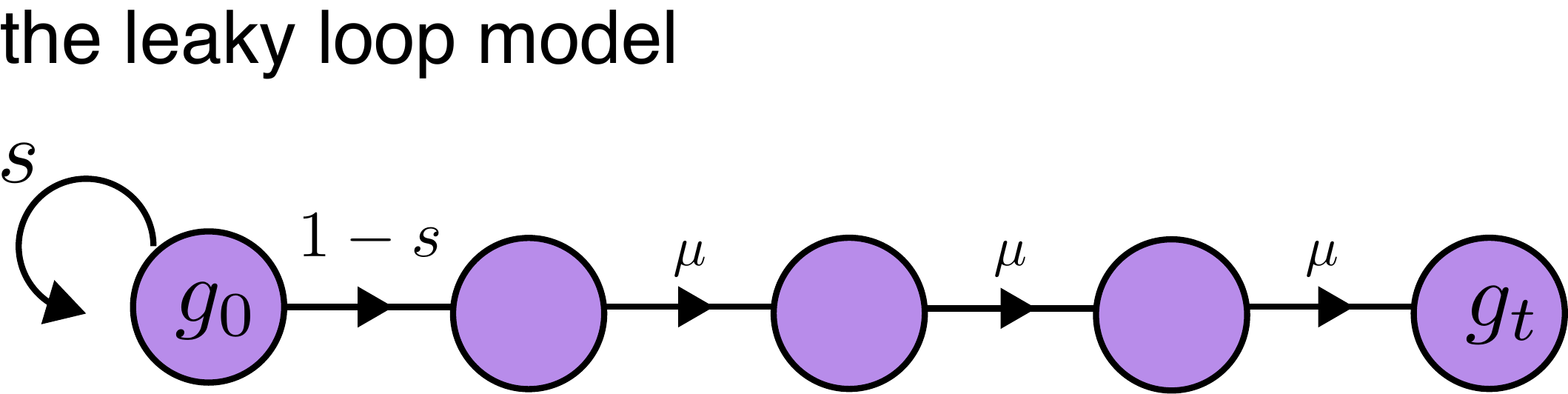}
    \caption{Schematic of the leaky loop model. See text for details.}
    \label{fig:lloop}
\end{figure}
To make this concrete, consider the simplest case where $\mathcal{H}$ consists of a single node $g_0$ with a self-loop. At each time step, the particle remains at $g_0$ with probability $s$ (the ``leakiness'' parameter) or enters the ballistic tail with probability $1-s$ (see Figure~\ref{fig:lloop}).

\begin{figure}
    \centering
    \includegraphics[width=\textwidth]{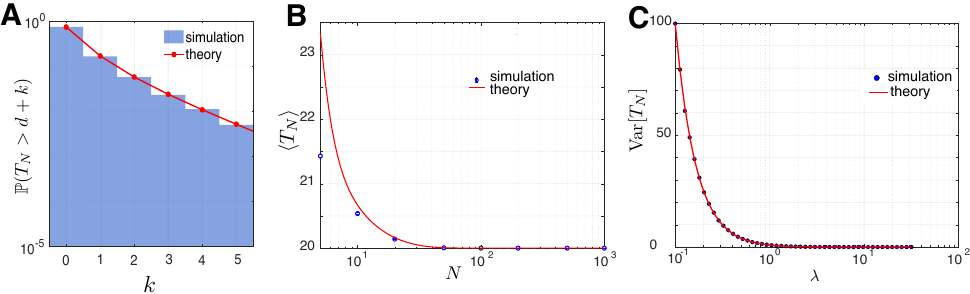}
    \caption{Comparison of Monte Carlo simulations and theory in the leaky loop model. (A) Comparison of the distribution generated from simulations with Eq.~\eqref{eq:distTN}; (B) Comparison of $\langle T_N\rangle$ from simulations with Eq.~\eqref{eq:meanTN2}; (C) Comparison of the variance in $T_N$ from simulations with variance derived from Eq.~\eqref{eq:moments}. Here, unless otherwise stated, $\lambda = 1$, $d = 20$, $\mu = 0.9$, and $s = 0.5$. We averaged over 50000 trials.}
    \label{fig:leakyloop}
\end{figure}

Here $d_{\mathcal{H}} = 1$. The shortest path probability is simply the probability of exiting immediately combined with the tail survival:
\begin{equation}
    p_d = (1-s)\mu^{d-1}.
\end{equation}
A delayed arrival of $d+k$ steps corresponds to the particle looping $k$ times before exiting. Thus:
\begin{equation}
    p_{d+k} = s^k(1-s)\mu^{d-1}.
\end{equation}
The cumulative probability ratio $F(k)$ is a geometric sum:
\begin{equation}
    F(k) = \sum_{j=0}^k \frac{s^j(1-s)\mu^{d-1}}{(1-s)\mu^{d-1}} = \frac{1-s^{k+1}}{1-s}.
\end{equation}
As $k \to \infty$, $F(k) \to (1-s)^{-1}$, a constant independent of $d$. Substituting this into our main result, the mean extreme first passage time for $N$ particles in this leaky loop geometry is:
\begin{equation}
    \langle T_N \rangle \approx d + \sum_{k=0}^\infty\exp\left(-\lambda(1-s)^{k+1}\right).
\end{equation}
This result highlights the ubiquity of the framework for processes dominated by injection noise: the delay is an additive constant determined by the source, not the path length.

\subsection{Failure of the Entropic Condition on the Bethe Lattice}

{To evaluate the limits of this framework on networks with exponential volume growth, we consider a sequence of triples on the Bethe lattice. Let $\mathcal{G}_n$ be an infinite Bethe lattice (regular tree) with coordination number $z \ge 3$. We fix the starting site $s_n = g_0$ for all $n$, and choose the target site $t_n$ such that it lies exactly at graph distance $d_n = n$ from $g_0$ along a specific, unique branch.} 

Because $\mathcal{G}_n$ is a tree, there exists a \emph{unique} geodesic connecting $g_0$ to $t_n$. We examine the probability ratio $p_{d+k}/p_d$ as the target is moved infinitely far away ($d \to \infty$).

The particle reaches the target in exactly $d$ steps if and only if it follows the unique geodesic at every step. At each step, there is exactly one neighbor moving toward the target out of $z$ choices:
\[
p_d = z^{-d}.
\]

Any path reaching the target in $d+2\ell$ steps must include $\ell$ ``wasteful'' excursions. Unlike the Comet graph sequence (where excursions were confined to the fixed head), on the Bethe lattice sequence, an excursion can be initiated from \emph{any} of the $d$ vertices along the geodesic.

\begin{figure}
    \centering
    \includegraphics[width=\textwidth]{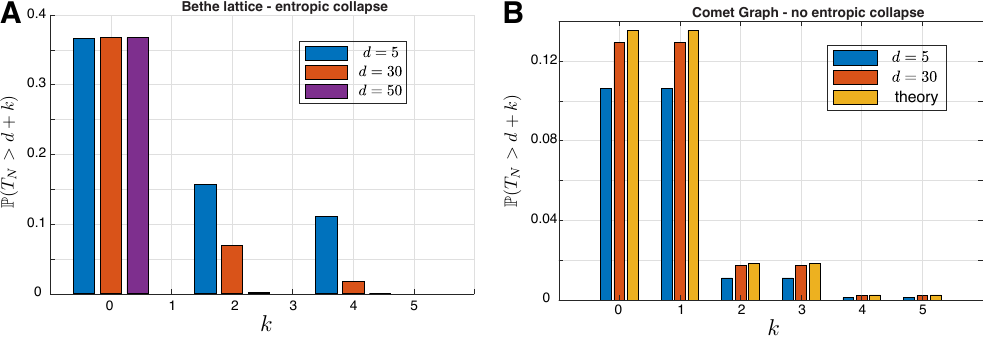}
    \caption{Entropic collapse in the Bethe lattice and lack thereof in the Comet graph. (A) In the Bethe lattice, the distribution of $T_N$ explicitly varies with $d$, causing our expression to fail. (B) The distribution is independent of $d$ in the Comet graph, yielding similar distributions for $T_N$ in the short $d$ and long $d$ cases, and providing evidence that entropic collapse is not impending. Here, $\lambda = 1$, $s = 0.5$, $N = 5000$, $z = 3$.}
    \label{fig:collapse}
\end{figure}

Consider the case of a single excursion ($\ell=1$, total steps $d+2$). The particle must traverse the $d$ geodesic edges and, at one intermediate vertex, step off the geodesic and immediately return. Note the following observations:

\begin{enumerate}
    \item There are approximately $d$ vertices along the path where this error can occur.
    \item At a specific vertex, the probability of stepping off the geodesic is $(z-2)/z$ (moving to a side branch).
    \item The probability of returning immediately is $1/z$.
\end{enumerate}

The probability of this specific trajectory scales as $p_{d+2} \propto d \cdot z^{-(d+2)}$. Consequently, the ratio depends linearly on $d$:
\[
\frac{p_{d+2}}{p_d} \propto \frac{d \cdot z^{-(d+2)}}{z^{-d}} \propto \frac{d}{z^2}.
\]
More generally, for $\ell$ excursions, the particle effectively chooses $\ell$ locations along a path of length $d$ to insert delays. Combinatorially, this scales as $\binom{d}{\ell} \approx d^\ell / \ell!$. 

\begin{figure}
    \centering
    \includegraphics[width=\textwidth]{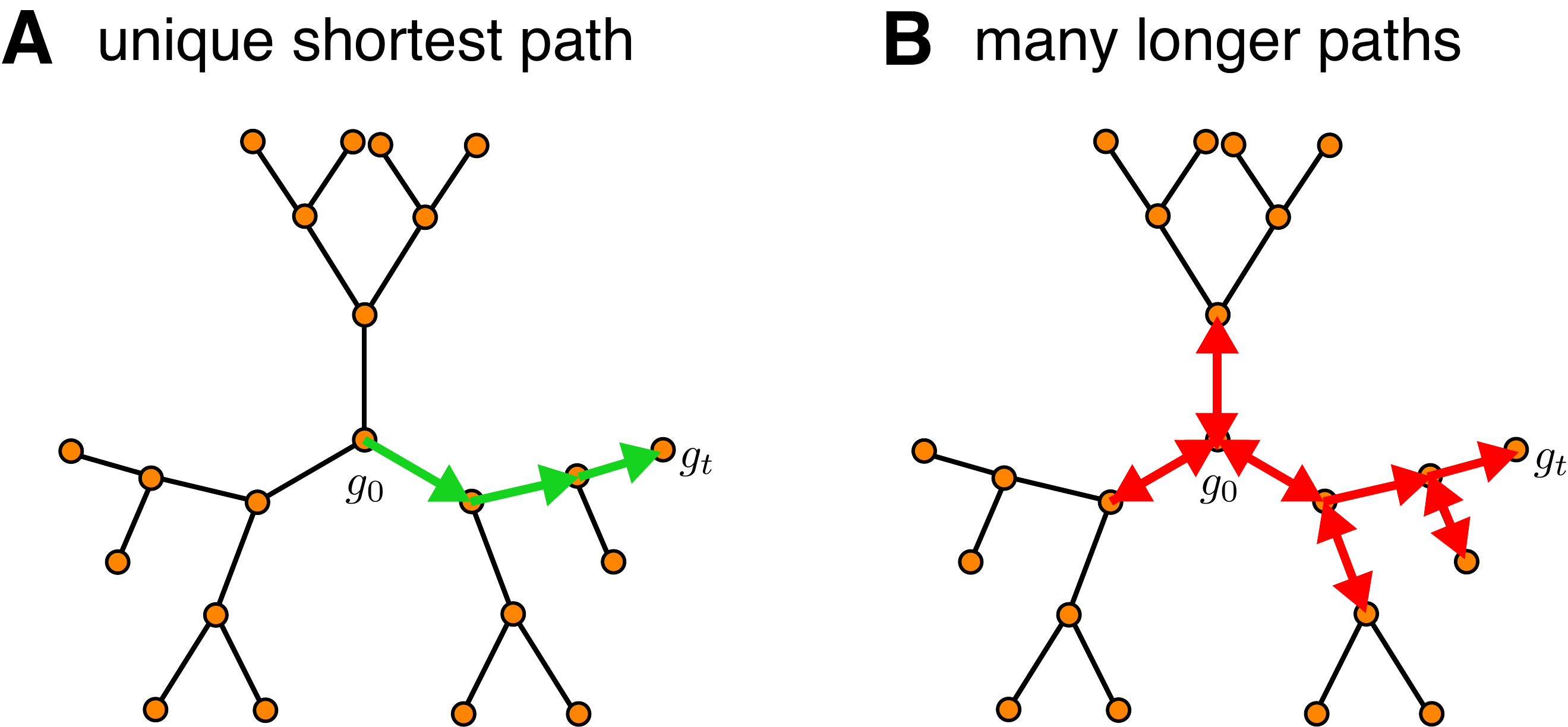}
    \caption{Entropic collapse in the Bethe lattice. (A) Schematic showing that there is a unique shortest path in a search on a Bethe lattice. (B) there are many more paths of length $d+k$. This underlies entropic collapse.}
    \label{fig:collapse_int}
\end{figure}

This leads to a breakdown of the entropic penalty condition. The cumulative ratio $F(k)$ is no longer constant:
\begin{equation}
    F(k) = \sum_{j=0}^k \frac{p_{d+j}}{p_d} \sim \sum_{\ell=0}^{\lfloor k/2 \rfloor} \frac{(C d)^\ell}{\ell!}.
\end{equation}
As $d \to \infty$, this ratio diverges. Thus, our entropy condition, Eq.~\eqref{eq:entropy}, is violated.

The key point here is that $F(k)$ explicitly depends on $d$ in the Bethe lattice example, causing divergence in the $d \to \infty$ limit. For our expressions to hold, the tail distribution must be \textit{independent} of $d$. In Figure~\ref{fig:collapse}, we compare the tail distributions $\mathbb{P}(T_N > d + k)$ in the Bethe lattice and the Comet graph for different $d$ values. In Figure~\ref{fig:collapse}A, we show that the distribution changes as $d$ is changed. For $d=5$, the distribution exhibits a longer tail, and the tail shortens as $d$ is increased. On the other hand, $\mathbb{P}(T_N > d + k)$ is independent of $d$ in the Comet graph (see Figure~\ref{fig:collapse}B). Our theoretical expressions agree with the distribution in the Comet graph for large or small $d$, as the distribution is independent of $d$.

In the Comet graph, the entropy is localized in the head. In the Bethe lattice, the entropy is distributed along the entire path. As $d$ increases, the number of ways to ``get lost'' grows, causing the probability mass to shift away from the minimum time $d$. 

Consequently, the distribution $p_t$ does not retain an exponential shape anchored at $t=d$. Instead, it \textbf{collapses} into a distribution dominated by discrete peaks at parities $d, d+2, ...,$ which, on a coarse-grained scale, look like a Gaussian centered at the mean time $\langle t \rangle = d/v_{drift}$ where $v_{drift} < 1$ is the effective ballistic velocity. This signifies a transition from the injection-limited class to the bulk-limited class. The failure of our entropic condition $F(k) \sim \text{constant}$ serves as a precise diagnostic for this phase transition. In other words, if $F(k)$ depends on $d$, then entropy is distributed along the entire graph and the particle dynamics represent a bulk-limited process. Conversely, if $F(k)$ is independent of $d$, the entropy is localized in the head and the particle dynamics represent an injection-limited process.

Next we look at an example of particle dynamics on a graph whose arrival probabilities are entropically bounded yet the extreme first passage distribution is multimodal, which is atypical for the classical extreme first passage distributions.

\subsection{The Braided Tail and the Tortoise-and-Hare Crossover}

{The failure of the entropic condition on the Bethe lattice raises a fundamental question about the topological boundary of Eq.~\eqref{eq:distTN}: does any deviation from a strictly 1D directed tail induce entropic collapse? To demonstrate that the tail can contribute non-trivially to the extreme value statistics without violating the entropic boundedness condition, we introduce a finite parallel-track geometry.}

{Suppose the tail $\mathcal{T}$ splits into $n$ parallel tracks at $g_{out}$, which merge at the target $g_t$. Each track $i \in \{1, \dots, n\}$ is chosen with probability $q_i$, such that $\sum_{i=1}^n q_i = 1$. Let Track 1 be the absolute fastest track, with length $L$. Other tracks $i$ have lengths $L + \Delta_i$ with $\Delta_i \ge 0$, representing detours. The survival probability along track $i$ is $\mu^{L+\Delta_i}$.}

{To evaluate the macroscopic limit for this topology, we define the sequence of triples such that the baseline length of the parallel tracks grows ($L_n = n \to \infty$), but the detours $\Delta_i$ and routing probabilities $q_i$ remain fixed.}

{The shortest possible distance to the target is $d = d_{\mathcal{H}} + L$. To arrive in exactly $d$ steps, the particle must execute a perfect run in the head and choose Track 1:
\begin{equation}
    p_d = \pi_{d_{\mathcal{H}}}(g_0, g_{out}) \cdot q_1 \cdot \mu^L.
\end{equation}
A delayed arrival of $d+k$ steps implies that if the particle took track $i$, it must have spent $k - \Delta_i$ additional steps in the head. Summing over the mutually exclusive track choices yields:
\begin{equation}
    p_{d+k} = \sum_{i=1}^n \pi_{d_{\mathcal{H}} + k - \Delta_i}(g_0, g_{out}) \cdot q_i \cdot \mu^{L+\Delta_i},
\end{equation}
where $\pi_m = 0$ for $m < d_{\mathcal{H}}$. Testing the entropic boundedness condition along our sequence, we evaluate the ratio:
\begin{equation}
    \frac{p_{d+k}}{p_d} = \sum_{i=1}^n \frac{\pi_{d_{\mathcal{H}} + k - \Delta_i}}{\pi_{d_{\mathcal{H}}}} \left(\frac{q_i}{q_1}\right) \mu^{\Delta_i}.
\end{equation}
Remarkably, the macroscopic distance $L$ completely cancels out of the sum. The entropic boundedness condition holds, and the extreme value distribution follows our exponential limit perfectly. Unlike the Bethe lattice, the entropy imparted by delayed paths is finite. Mathematically, $F(k)$ operates as a convolution: the total delay a particle experiences is simply the sum of two independent events. It is the time the particle spends trapped in the source graph, combined with the extra distance it travels if it is routed onto a slower parallel track.} 

\begin{figure}
    \centering
    \includegraphics[width=\textwidth]{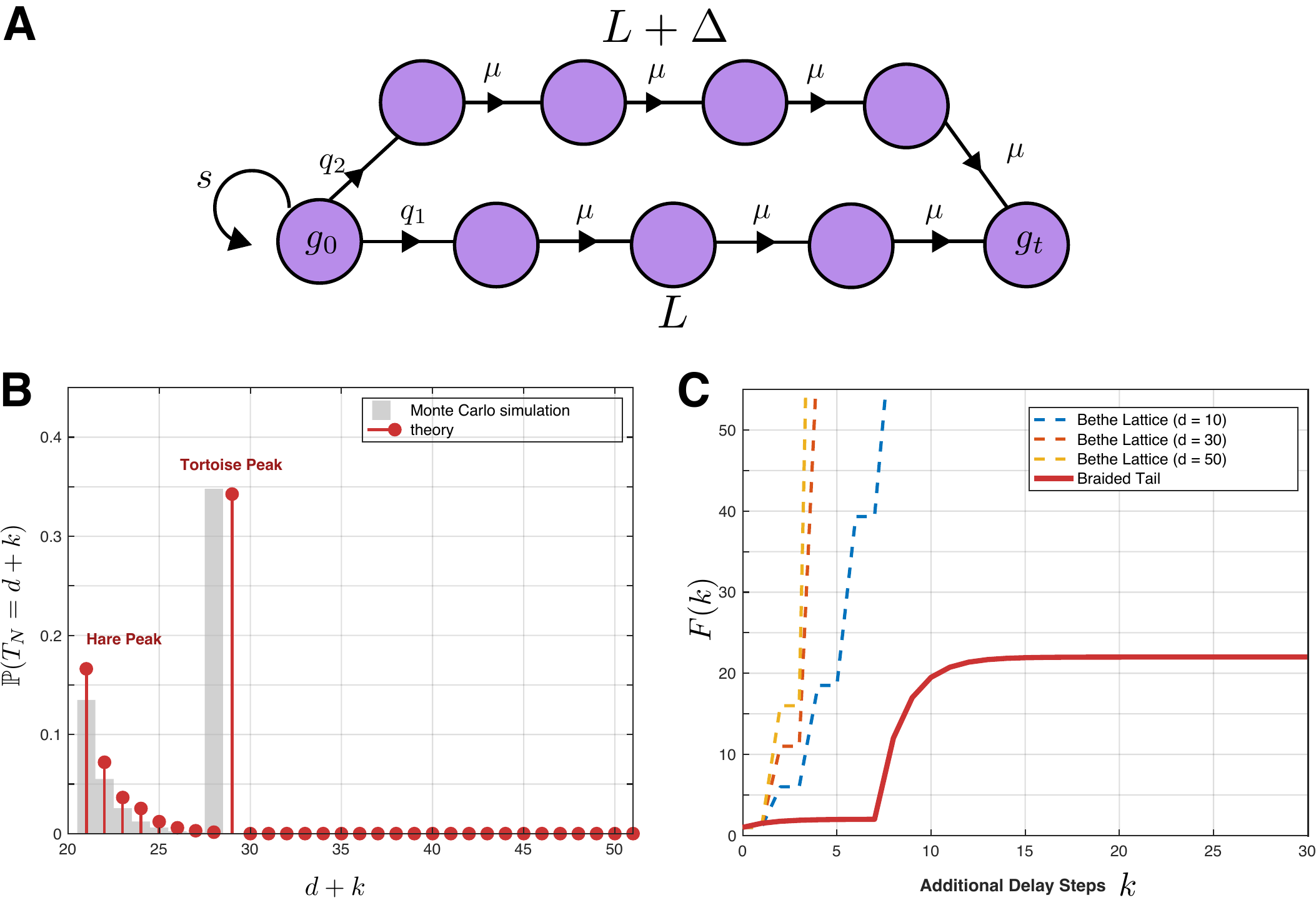}
    \caption{Results for the braided tail. (A) Schematic of the two lane braided tail analyzed in the text. (B) The Tortoise and the Hare regime arises when $(q_2/q_1)\mu^\Delta \gg 1$ and yields a multimodal distribution for $T_N$. Here, $s = 0.5$, $\mu = 0.95$, $L = 20$, $\Delta = 8$, $q_1 = 1E-4$, and $q_2 = 0.9999$. (C) Plots of $F(k)$ for the braided tail and the Bethe lattice, showing the entropic boundedness of the braided tail regardless of $d$ and the divergence of $F(k)$ for the Bethe lattice as $d$ varies.}
    \label{fig:collapse-vs-boundedness}
\end{figure}

{This finite multi-track topology can yield strongly multimodal extreme value distributions. Consider a two-track ``Tortoise and Hare" regime. {In this case, the entropic penalty function $F(k)$ has the piecewise structure
\begin{align*}
F(k) &= \sum_{j=0}^{k} \frac{p_{L+j}}{p_L} \\
&= 
\begin{cases} 
\displaystyle \sum_{j=0}^{k} s^j = \frac{1-s^{k+1}}{1-s}, & k < \Delta \\
\\
\displaystyle \sum_{j=0}^{k} s^j + \frac{q_2}{q_1}\mu^\Delta \sum_{j=\Delta}^{k} s^{j-\Delta} = \frac{1-s^{k+1}}{1-s} + \frac{q_2}{q_1}\mu^\Delta \frac{1-s^{k-\Delta+1}}{1-s}, & k \ge \Delta 
\end{cases}
\end{align*}}  
{The function fragments at $k = \Delta$ because any delay less than $\Delta$ must have occurred due to the particle being stuck in the head. A delay of $\Delta$ or greater can be caused by a particle being stuck in the head or taking the longer path (see Figure~\ref{fig:collapse-vs-boundedness}A). Regardless of the path, $F(k)$ remains independent of $d$, suggesting that the paths affiliated with the graph are entropically bounded.}

{To see when a multimodal distribution can arise, consider the following regime:} Track 1 (the Hare) is fast ($\Delta_1=0$) but rarely chosen ($q_1 \ll 1$), while Track 2 (the Tortoise) is slower ($\Delta_2 = \Delta > 0$) but highly probable ($q_2 \approx 1$) (see Figure~\ref{fig:collapse-vs-boundedness}A). If the routing volume overwhelms the degradation such that $(q_2/q_1)\mu^\Delta \gg 1$, the step-size $p_{d+k}/p_d$ becomes starkly non-monotone. This fractures the extreme value distribution, creating a primary peak at $d$ generated by rare fast arrivals, and a massive secondary peak at $d+\Delta$ generated by the bulk routing. This crossover demonstrates that discrete macroscopic tails can act as highly complex, dynamic geometric filters without triggering entropic collapse (see Figure~\ref{fig:collapse-vs-boundedness}B-C).}

In Figure ~\ref{fig:collapse-vs-boundedness}C, we again illustrate our point about the relationship between $F(k)$ and entropic boundedness. Regardless of the complexity and length of the tail linking $g_{out}$ to $g_t$, the shape of $F(k)$ is independent of $d$, which indicates that the braided tail, even with a large $n$ number of lanes, is entropically bounded. On the other hand, $F(k)$ diverges as $d$ grows along the Bethe lattice, indicating a lack of entropic boundedness.

\section{Discussion}

This work establishes an extreme-value scaling theory for first-passage processes on discrete hierarchical graphs. In contrast to classical extreme-value theory for continuum stochastic processes, where the support of the first-passage distribution extends continuously to zero, discrete-time random walks exhibit a lower bound that accumulates nontrivial probability mass at the graph distance between the source and the target. {Previous work has established how minimal time bounds shape extreme-value frameworks in specific networks and diffusion contexts~\cite{lawley2020extreme,linn2022extreme}, but our work explicitly describes the structure of extreme-value times in the presence of minimal bounds imputed by the topology of the graph.} This structural difference fundamentally alters the nature of the extreme statistics. {Indeed, our work here points to, but does not prove, regimes where the continuum limit coincides with the discrete process.}

Our main result shows that, in a broad class of discrete geometries, the earliest arrival among many independent walkers is governed by a Poissonian competition between rare, ballistic trajectories. The resulting extreme-value distribution is neither Gumbel, Fr\'{e}chet, nor Weibull, but instead takes a exponential form controlled by a geometry-dependent function $F(k)$, which encodes the entropic suppression of detours away from shortest paths.

An important takeaway from this work is that in discrete hierarchical networks as discussed in this paper, the length and complexity of the tail does not impact the shape of $F(k)$. We point out that this is not a trivial result emanating from unidirectional dynamics along the tail. At first glance, one might assume that for entirely directed, unidirectional tail geometries, the preservation of the source delay distribution is a trivial consequence of the lack of bulk routing choices. However, this intuition fails in the extreme-value limit. As the macroscopic distance $d \to \infty$, the probability of a particle traversing the tail without degradation decays exponentially. It is highly non-trivial that this severe, exponential filtering by the bulk acts strictly as a scalar modulation $(\lambda)$, rather than warping the shape of the delay distribution $(F(k))$. Our main result formally proves that, provided the network remains entropically bounded, the signature of the localized source is invariant to the macroscopic length of the transport channel.

\textit{Scope and limitations.} A central assumption of our framework is that the cumulative probability of arrivals occurring within a fixed number of steps beyond the graph distance remains asymptotically proportional to the shortest-path probability. This assumption captures the idea that while longer paths proliferate combinatorially, their total probability weight is suppressed by an entropic penalty. When this balance holds, the extreme statistics are dominated by near-geodesic trajectories, and the scaling limit described in Eq.~\eqref{eq:meanTN} applies.

 We also assume equal-probability random walks within $\mathcal{H}$; the extension to biased or heterogeneous transition rates, relevant for systems such as kinesin stepping and router queuing, remains an open direction.

Importantly, not all discrete geometries satisfy this condition. In particular, graphs with exponential volume growth along geodesics—such as infinite regular trees—exhibit a proliferation of near-geodesic excursions whose cumulative probability grows with the source–target distance. In such cases, the present scaling regime breaks down, and the extreme statistics fall into a different class. Rather than representing a failure of the theory, this delineates a sharp boundary between ballistic-dominated and entropy-dominated search processes. This boundary is precisely captured by the entropic boundedness condition introduced in Section~\ref{sec:framework}. We summarize this delineation as follows: if $F(k)$ is independent of $d$, then our Eq.~\eqref{eq:distTN} holds.

This distinction highlights a key message of the paper: the geometry of the underlying graph directly determines the  class of extreme first-passage behavior.

\textit{Physical interpretation.} From a physical perspective, the extreme-event regime analyzed here corresponds to a competition between many independent searchers, each attempting to realize an atypically efficient trajectory. The earliest arrival is not produced by diffusive exploration but by rare, nearly deterministic paths that minimize distance while tolerating only a limited number of local fluctuations. This, again, is a differentiation between injection-limited processes and bulk-limited processes, for which our entropy condition provides precise thresholds.

The function $F(k)$ provides a quantitative measure of how quickly this ballistic regime deteriorates as additional steps are allowed. In geometries where $F(k)$ remains finite for fixed $k$, the search process admits a well-defined extreme-value scaling limit in which the first arrival behaves as a Poisson process of optimal trajectories. In geometries where $F(k)$ grows with system size, the dominance of shortest paths is lost, signaling a crossover to a fundamentally different search mechanism.

\textit{Relation to previous work.}  {Previous studies of extreme first-passage times have focused primarily on continuum diffusion, branching processes, or mean-field approximations, where shortest paths are not sharply distinguished from nearby trajectories~\cite{lawley2020extreme,linn2022extreme}. Recent work has also investigated discrete-time random walks in randomly fluctuating environments, demonstrating that environmental disorder drives the variance of extreme arrivals to follow universal power laws~\cite{hass2024extreme}. Our work addresses a complementary question: what structural property of a deterministic graph determines whether shortest-path arrivals continue to govern extreme statistics in the large-distance limit? By isolating the role of graph topology rather than environmental disorder, the present work emphasizes that lattice structure, neighborhood rules, and graph growth properties play a decisive role in shaping extreme statistics.}

Our results complement recent interest in fastest-particle phenomena by identifying a class of systems where the fastest arrival is controlled by geometric constraints rather than by large deviations of Brownian motion. This perspective is particularly relevant for applications involving discrete transport networks, constrained search algorithms, and biological or technological systems where motion occurs in quantized steps~\cite{kolomeisky2007molecular,dora2020active,bressloff2016model,zhang1989new,zhang1995service}.

\textit{Outlook.} Several directions emerge naturally from this work. One is the systematic classification of graphs according to whether they satisfy the entropic constraint underlying Eq.~\eqref{eq:distTN}, thereby mapping out  classes of discrete extreme-value behavior. Another is the extension of the framework to graphs with weak inhomogeneities or finite-size effects, where crossover phenomena between ballistic and entropy-dominated regimes may arise.

More broadly, this study underscores the importance of discrete geometry in extreme-event statistics. While continuum limits provide powerful approximations, they can obscure qualitative features that emerge only when the underlying discreteness of space and time is retained. Understanding these features is essential for accurately modeling extreme outcomes in realistic search and transport processes.

\section*{Acknowledgements}
 BRK would like to thank his wife, Hajra Habib, and two sons, Surya and Taraq, for being the lights of his life.

\vspace{10pt}
\bibliographystyle{unsrt}

\begin{thebibliography}{10}

\bibitem{chou2014first}
Tom Chou and Maria~R D'Orsogna.
\newblock First passage problems in biology.
\newblock In {\em First-passage phenomena and their applications}, pages
  306--345. World Scientific, 2014.

\bibitem{patie2004some}
Pierre Patie.
\newblock {\em On some first passage time problems motivated by financial
  applications}.
\newblock PhD thesis, Universit{\"a}t Z{\"u}rich, 2004.

\bibitem{redner2001guide}
Sidney Redner.
\newblock {\em A guide to first-passage processes}.
\newblock Cambridge university press, 2001.

\bibitem{hillen2025mean}
Thomas Hillen, Maria~R D’Orsogna, Jacob~C Mantooth, and Alan~E Lindsay.
\newblock Mean first passage times for transport equations.
\newblock {\em SIAM Journal on Applied Mathematics}, 85(1):78--108, 2025.

\bibitem{d2026mean}
Maria~R D’Orsogna, Alan~E Lindsay, and Thomas Hillen.
\newblock Mean first passage times of higher-dimensional velocity jump
  processes.
\newblock {\em Physical Review Letters}, 136(24):247102, 2026.

\bibitem{newby2010quasi}
Jay~M Newby and Paul~C Bressloff.
\newblock Quasi-steady state reduction of molecular motor-based models of
  directed intermittent search.
\newblock {\em Bulletin of mathematical biology}, 72(7):1840--1866, 2010.

\bibitem{kolomeisky2007molecular}
Anatoly~B Kolomeisky and Michael~E Fisher.
\newblock Molecular motors: a theorist's perspective.
\newblock {\em Annu. Rev. Phys. Chem.}, 58(1):675--695, 2007.

\bibitem{sirovich2011spiking}
Lawrence Sirovich and Bruce Knight.
\newblock Spiking neurons and the first passage problem.
\newblock {\em Neural computation}, 23(7):1675--1703, 2011.

\bibitem{byrne2012using}
Michael~E Byrne and Michael~J Chamberlain.
\newblock Using first-passage time to link behaviour and habitat in foraging
  paths of a terrestrial predator, the racoon.
\newblock {\em Animal Behaviour}, 84(3):593--601, 2012.

\bibitem{masuda2017random}
Naoki Masuda, Mason~A Porter, and Renaud Lambiotte.
\newblock Random walks and diffusion on networks.
\newblock {\em Physics reports}, 716:1--58, 2017.

\bibitem{noh2004random}
Jae~Dong Noh and Heiko Rieger.
\newblock Random walks on complex networks.
\newblock {\em Physical review letters}, 92(11):118701, 2004.

\bibitem{saacke2000relationship}
RG~Saacke, JC~Dalton, S~Nadir, RL~Nebel, and JH~Bame.
\newblock Relationship of seminal traits and insemination time to fertilization
  rate and embryo quality.
\newblock {\em Animal reproduction science}, 60:663--677, 2000.

\bibitem{kumari2024first}
Suman Kumari, Partha~Sarathi Mandal, and Moitri Sen.
\newblock First passage time and peak size probability distributions for a
  complex epidemic model.
\newblock {\em The European Physical Journal Plus}, 139(4):1--20, 2024.

\bibitem{dora2020active}
Matteo Dora and David Holcman.
\newblock Active flow network generates molecular transport by packets: case of
  the endoplasmic reticulum.
\newblock {\em Proceedings of the Royal Society B}, 287(1930):20200493, 2020.

\bibitem{schuss2019redundancy}
Z~Schuss, K~Basnayake, and D~Holcman.
\newblock Redundancy principle and the role of extreme statistics in molecular
  and cellular biology.
\newblock {\em Physics of life reviews}, 28:52--79, 2019.

\bibitem{patra2021level}
Swayamshree Patra and Debashish Chowdhury.
\newblock Level crossing statistics in a biologically motivated model of a long
  dynamic protrusion: passage times, random and extreme excursions.
\newblock {\em Journal of Statistical Mechanics: Theory and Experiment},
  2021(8):083207, 2021.

\bibitem{basnayake2019fast}
Kanishka Basnayake, David Mazaud, Alexis Bemelmans, Nathalie Rouach, Eduard
  Korkotian, and David Holcman.
\newblock Fast calcium transients in dendritic spines driven by extreme
  statistics.
\newblock {\em PLoS biology}, 17(6):e2006202, 2019.

\bibitem{wong2026first}
Tony Wong, Ikchang Cho, Maria~R D’Orsogna, and Tom Chou.
\newblock First passage times to t cell activation.
\newblock {\em SIAM Journal on Life Sciences}, 1(2):229--261, 2026.

\bibitem{iyer2016first}
Srividya Iyer-Biswas and Anton Zilman.
\newblock First-passage processes in cellular biology.
\newblock {\em Advances in chemical physics}, 160:261--306, 2016.

\bibitem{allard2026tethered}
Jun Allard and Omer Dushek.
\newblock Tethered signaling proteins.
\newblock {\em Annual Review of Biophysics}, 55, 2026.

\bibitem{karamched2020bayesian}
Bhargav Karamched, Simon Stolarczyk, Zachary~P Kilpatrick, and Kresimir Josic.
\newblock Bayesian evidence accumulation on social networks.
\newblock {\em SIAM journal on applied dynamical systems}, 19(3):1884--1919,
  2020.

\bibitem{karamched2020heterogeneity}
Bhargav Karamched, Megan Stickler, William Ott, Benjamin Lindner, Zachary~P
  Kilpatrick, and Kre{\v{s}}imir Josi{\'c}.
\newblock Heterogeneity improves speed and accuracy in social networks.
\newblock {\em Physical review letters}, 125(21):218302, 2020.

\bibitem{stickler2023impact}
Megan Stickler, William Ott, Zachary~P Kilpatrick, Kre{\v{s}}imir Josi{\'c},
  and Bhargav~R Karamched.
\newblock Impact of correlated information on pioneering decisions.
\newblock {\em Physical review research}, 5(3):033020, 2023.

\bibitem{linn2024fast}
Samantha Linn, Sean~D Lawley, Bhargav~R Karamched, Zachary~P Kilpatrick, and
  Kre{\v{s}}imir Josi{\'c}.
\newblock Fast decisions reflect biases; slow decisions do not.
\newblock {\em Physical Review E}, 110(2):024305, 2024.

\bibitem{lawley2020distribution}
Sean~D Lawley.
\newblock Distribution of extreme first passage times of diffusion.
\newblock {\em Journal of Mathematical Biology}, 80(7):2301--2325, 2020.

\bibitem{lawley2020extreme}
Sean~D. Lawley.
\newblock Extreme first-passage times for random walks on networks.
\newblock {\em Phys. Rev. E}, 102:062118, Dec 2020.

\bibitem{lawley2020probabilistic}
Sean~D Lawley and Jacob~B Madrid.
\newblock A probabilistic approach to extreme statistics of brownian escape
  times in dimensions 1, 2, and 3.
\newblock {\em Journal of Nonlinear Science}, 30(3):1207--1227, 2020.

\bibitem{lawley2020universal}
Sean~D Lawley.
\newblock Universal formula for extreme first passage statistics of diffusion.
\newblock {\em Physical Review E}, 101(1):012413, 2020.

\bibitem{linn2022extreme}
Samantha Linn and Sean~D Lawley.
\newblock Extreme hitting probabilities for diffusion.
\newblock {\em Journal of Physics A: Mathematical and Theoretical},
  55(34):345002, 2022.

\bibitem{lawley2021extreme}
Sean~D Lawley.
\newblock Extreme first passage times of piecewise deterministic markov
  processes.
\newblock {\em Nonlinearity}, 34(5):2750, 2021.

\bibitem{maclaurin2025extreme}
James MacLaurin and Jay Newby.
\newblock Extreme first passage times for populations of identical rare events.
\newblock {\em SIAM Journal on Applied Mathematics}, 85(1):109--142, 2025.

\bibitem{tauber2014critical}
Uwe~C T{\"a}uber.
\newblock {\em Critical dynamics: a field theory approach to equilibrium and
  non-equilibrium scaling behavior}.
\newblock Cambridge University Press, 2014.

\bibitem{bressloff2016model}
Paul~C Bressloff and Bhargav~R Karamched.
\newblock Model of reversible vesicular transport with exclusion.
\newblock {\em Journal of Physics A: Mathematical and Theoretical},
  49(34):345602, 2016.

\bibitem{gramm2004automated}
Jens Gramm, Jiong Guo, Falk H{\"u}ffner, and Rolf Niedermeier.
\newblock Automated generation of search tree algorithms for hard graph
  modification problems.
\newblock {\em Algorithmica}, 39(4):321--347, 2004.

\bibitem{hass2024extreme}
Jacob~B Hass, Hindy Drillick, Ivan Corwin, and Eric~I Corwin.
\newblock Extreme diffusion measures statistical fluctuations of the
  environment.
\newblock {\em Physical Review Letters}, 133(26):267102, 2024.

\bibitem{bressloff2013stochastic}
Paul~C Bressloff and Jay~M Newby.
\newblock Stochastic models of intracellular transport.
\newblock {\em Reviews of Modern Physics}, 85(1):135--196, 2013.

\bibitem{van1992stochastic}
NG~Van~Kampen.
\newblock Stochastic processes in physics and chemistry.
\newblock {\em (No Title)}, 1992.

\bibitem{gardiner2009stochastic}
Crispin Gardiner.
\newblock {\em Stochastic methods}, volume~4.
\newblock Springer Berlin Heidelberg, 2009.

\bibitem{flajolet2006ubiquitous}
Philippe Flajolet.
\newblock The ubiquitous digital tree.
\newblock In {\em Annual Symposium on Theoretical Aspects of Computer Science},
  pages 1--22. Springer, 2006.

\bibitem{zhang1989new}
Lixia Zhang.
\newblock A new architecture for packet switching network protocols.
\newblock Technical report, 1989.

\bibitem{zhang1995service}
Hui Zhang.
\newblock Service disciplines for guaranteed performance service in
  packet-switching networks.
\newblock {\em Proceedings of the IEEE}, 83(10):1374--1396, 1995.

\end{thebibliography}

\end{document}